\documentclass[11pt]{article}
\usepackage{amssymb,amstext,amsmath,amsthm}
\usepackage[dvips]{graphicx}
\usepackage{latexsym}
\usepackage{psfrag}
\usepackage{amsfonts}
\usepackage{bbm}
\usepackage{color}

\newcommand{\beqa}{\begin{eqnarray}}
\newcommand{\eeqa}{\end{eqnarray}}
\newcommand{\f}{\begin{equation}}
\newcommand{\ff}{\end{equation}}

\newcommand{\bean}{\begin{eqnarray*}}
\newcommand{\eean}{\end{eqnarray*}}
\newcommand{\ra}{\rightarrow}

\newcommand{\pa}{\partial}

\def\be{\begin{equation}} \def\ee{\end{equation}}

\begin{document}

\title{\Large\bf Static Isotropic Spacetimes with Radially Imperfect
Fluids}
\author{{Tomasz Konopka}\\
{\small ${}$\emph{ITP, Utrecht University, Utrecht 3584 CE, The
Netherlands}}}
\date{}
\maketitle
\begin{abstract}
When solving the equations of General Relativity in a symmetric
sector, it is natural to consider the same symmetry for the
geometry and stress-energy. This implies that for static and
isotropic spacetimes, the most general natural stress-energy
tensor is a sum of a perfect fluid and a radial imperfect fluid
component. In the special situations where the perfect fluid
component vanishes or is a spacetime constant, the solutions to
Einstein's equations can be thought of as modified Schwarzschild
and Schwarzschild-de Sitter spaces. Exact solutions of this type
are derived and it is shown that whereas deviations from the
unmodified solutions can be made small, among the manifestations
of the imperfect fluid component is a shift in angular momentum
scaling for orbiting test-bodies at large radius. Based on this
effect, the question of whether the imperfect fluid component can
feasibly describe dark matter phenomenology is addressed.
\end{abstract}

\section{Introduction}

The observed Universe is homogenous at very large scales but not
at small ones where it contains numerous localized matter sources.
Within the well-tested \cite{WillTest} theory of General
Relativity \cite{Waldbook}, this means the geometry on large
scales can be described by a metric of the
Friedman-Robertson-Walker type while at smaller scales it must be
approximated by metrics of the Schwarzschild or Schwarzschild-de
Sitter forms. These idealized metrics can be derived from the
equations of motion of General Relativity by assuming different
symmetries at the different scales.

It may be argued that when solving the equations of motion of
General Relativity in some symmetric sector, it is natural to
consider the same symmetry for the geometry $g_{\mu\nu}$ and
stress-energy $T_{\mu\nu}$ to which the geometry couples.
Accordingly, when assuming spatial homogeneity for the geometry,
it is natural to take $T_{\mu\nu}$ to contain one free function
associated with the time direction and one free function
associated with the spatial directions. A stress-tensor of this
kind is said to be in perfect fluid form and can be written as \be
\label{Tpf} T^{pf}_{\mu\nu} = (\rho+p)U_\mu U_\nu + p \,g_{\mu\nu}
\ee with $U^\mu$ a unit time-like vector, and $\rho$ and $p$ some
functions called the energy density and pressure, respectively. In
contrast, for static isotropic geometries, what is natural for
$T_{\mu\nu}$ is that it should have three independent functions -
two associated with the time and radial directions, and one
associated with the remaining two angular directions. A stress
tensor of this type can be written as \be \label{T_CMO} T_{\mu\nu}
= T_{\mu\nu}^{pf} + q V_{\mu}V_{\nu} \ee where the first term on
the right hand side is of perfect fluid form (\ref{Tpf}), and in
the second term $V^\mu$ is a unit vector pointing in the radial
direction and $q$ is some function. Since the modification to the
perfect fluid form comes only in the radial components of
$T_{\mu\nu}$, a stress-energy tensor of this type may be called
radially imperfect. In other terminology it may be said to contain
viscous shear. It is the most general stress-energy tensor
compatible with staticity and isotropy \cite{Mannheim:2005bfa}.

In principle, the stress-energy tensor can be calculated in any
theory of matter coupled to gravity given a description of the
matter distribution. Recently, Cox, Mannheim, and O'Brien
performed calculations of this kind assuming a matter component in
the form of an incoherent collection of modes of a free
(minimally-coupled) massless scalar field \cite{CMO}. Reproducing
earlier results \cite{Mannheim:1985pp}, they showed that such a
matter distribution puts the stress-energy tensor into the perfect
fluid form when the background spacetime is Minkowski. Crucially,
however, they also showed that the incoherent collection of field
modes does not necessarily generate a perfect fluid stress-energy
tensor if the background spacetime is nonhomogenous. In one
analytic calculation involving a special static isotropic
background, they put the stress-energy tensor into the form
(\ref{T_CMO}) and explicitly wrote the functional form for $\rho$,
$p$ and $q$ with $q\neq 0$. This corroborates the motivation for
using the general form (\ref{T_CMO}) for the stress-tensor source
when it is coupled to a static and isotropic geometry.

The calculations in \cite{CMO} were done in the spirit of field
theory on curved spacetime. That is, the background spacetime was
assumed to be fixed. The approach was useful because it allowed
those authors to determine the field modes explicitly, sum over
them, and thus obtain the stress-energy tensor. From the
nonperturbative point of view, however, it is important to take
into account backreaction, i.e. to obtain a consistent solution to
Einstein equations with an imperfect fluid source. The
observations in \cite{CMO} are a direct motivation for doing
exactly that.

The purpose of the present work is to study static and isotropic
spacetimes coupled with the imperfect fluid stress-energy tensor
(\ref{T_CMO}). A similar task has been attempted in
\cite{Mannheim:2005bfa} but the issue is revisited here in more
detail. Attention is focused on describing spacetimes in which the
imperfect fluid component is nonvanishing and comparing them to
known solutions, like Schwarzschild, in which this component is
exactly zero.

Section \ref{s_static} deals with Einstein's equations for static
and isotropic spacetimes coupled to certain fluids of the form
(\ref{T_CMO}). Although isotropy may in some systems also be
thought to be an emergent property due to matter coupling
\cite{Konopka:2009vq, Konopka:2009yx}, in this work it is imposed
strongly. Several solutions -- modifications of the Schwarzschild,
anti-de Sitter, and Schwarzschild-de Sitter spaces -- are derived
and their properties are discussed. Then, some phenomenological
aspects of the modified black-hole solution are studied in Sec.
\ref{s_pheno}. Since one aspect of the phenomenology is similar to
that attributed to dark-matter in galaxies, Sec.
\ref{s_discussion} explores whether the imperfect fluid component
can actually be considered a candidate description for dark
matter. A brief summary of the results appears in
Sec.\ref{s_discussion}.

\section{Static Isotropic Solutions \label{s_static}}

In units where the speed of light and the Newton constant are set
to one, the equations of motion of General Relativity are \be
G_{\mu\nu} =8\pi T_{\mu\nu} \ee with $G_{\mu\nu}$ the Einstein
tensor derived from a metric $g_{\mu\nu}$, and $T_{\mu\nu}$ the
stress-energy tensor of matter \cite{Waldbook}. This section deals
with solutions to these equations under the assumption that the
geometry is static and isotropic (spherically-symmetric) and the
stress tensor is in the radially-imperfect form (\ref{T_CMO}).

With the assumed symmetries, the line element $ds^2 = g_{\mu\nu}
\, dx^\mu\, dx^\nu $ associated with a metric $g_{\mu\nu}$ in
coordinates $x^\mu$ can be written as \be \label{ds2_sc} ds^2 =
-f^2\, dt^2 + h^{-1}\, dr^2 + r^2 \,d\Omega^2 \ee so that the only
unfixed components of the metric are $g_{tt}=-f^2$ and
$g_{rr}=h^{-1}$. The functions $f$ and $h$ depend on the radial
coordinate $r$ only and $d\Omega^2 = d\theta^2+\sin^2\theta\,
d\phi^2$ is the line element of a unit sphere. With this choice of
coordinates and parametrization, the nonzero components of the
Einstein tensor occur only on the diagonal and are
\begin{subequations} \label{Gcomponents_sc}
\begin{align}
G_{tt} &= \frac{f^2}{r^2}\left(1-h-rh^\prime \right) \\
G_{rr} &=  \frac{1}{r^2hf} \left(hf-f+2rhf^\prime\right) \\
G_{\theta\theta} &= \frac{r}{2f} \left( 2h\left( f^\prime +
rf^{\prime\prime} \right)+h^\prime \left( f+rf^\prime \right)
\right) \\
G_{\phi\phi} &= \sin^2 \theta \, G_{\theta\theta}.
\end{align}
\end{subequations}

With a radially imperfect fluid of the form (\ref{T_CMO}), the
equations of motion become
\begin{subequations} \label{Gall_sc}
\begin{align}
G_{tt} &= \rho \, f^2, \label{Gtt_sc} \\
G_{rr} &= (p+q) \, h^{-1}, \label{Grr_sc} \\
G_{\theta\theta}  &= p\, r^2, \label{Gaa_sc} \\
G_{\phi\phi} &= p\, r^2 \sin^2\theta, \label{Ga2a2}
\end{align} \end{subequations}
where $\rho,$ $p$ and $q$ are functions only of $r$, and factors
of $8\pi$ have been absorbed on the right hand side. Since
equations (\ref{Ga2a2}) and (\ref{Gaa_sc}) are multiples of each
other, there are actually only three independent equations. And
since there are five unknown functions, the equations cannot be
solved without specifying additional data. If the geometry, i.e.
the functions $f$ and $h$, is specified then the equations can be
solved directly to yield the compatible $\rho$, $p$ and $q$. In
this way, any spacetime whatsoever can be obtained by some choices
for the stress-energy tensor. But it is more constructive to
specify reasonable sources, i.e. $\rho$, $p$ and $q$, and
subsequently solve for the compatible geometry. Below are a few
cases characterized by idealized choices for the energy density
$\rho$ and pressure $p$.

\subsection{Exactly Vanishing Perfect Fluid \label{s_newhole}}

The case of the vanishing perfect fluid is defined by \be \rho=p=0
\ee everywhere. In this case, equation (\ref{Gtt_sc}) can first be
solved for $h$. Next, the solution for $h$ can be inserted into
(\ref{Gaa_sc}) to yield a linear equation for $f$. The results are
\begin{align} \label{hbhsol}
h &= 1 - \frac{b}{r} \\ \label{fbhsol} f &= a
\left(\sqrt{1-\frac{b}{r}}\right)-c\left( 1 - \sqrt{1-\frac{b}{r}}
\, \mathrm{log} \left(\frac{\sqrt{r}+\sqrt{r-b}}{\sqrt{d}}
\right)\right) \end{align} where $a,b,c,d$ are constants. $a$ and
$c$ are dimensionless, while $b$ and $d$ have units of length.
(The constant $c$ should not be confused with the speed of light
which is set to one throughout by a choice of units.) These
functions $f$ and $h$ determine the metric. When they are inserted
into (\ref{Grr_sc}), they determine the imperfect fluid component
to be \be q = \frac{c}{r^2 f}. \label{qbh} \ee

If the imperfect fluid component is required to vanish, then the
constant $c$ must be set to zero. When $h$ in (\ref{hbhsol}) and
$f$ in (\ref{fbhsol}) with $c=0$ are inserted into (\ref{ds2_sc}),
they yield the Schwarzschild line-element with an additional
numeric constant $a^2$ in front of the $dt^2$ component. Because
of reparametrization invariance, this constant can be set to unity
without loss of generality. Thus the standard black hole solution,
with a horizon at the Schwarzschild radius $r=b$ and a singularity
at $r=0$, is recovered.

\begin{figure}[tbp]
\begin{center}
\begin{picture}(0,0)(0,0)
  \put(358,12){$\log_{10}r/b$}
  \put(38,127){$f^2$}
  \put(285,59){\small $c\!=\!0$}
  \put(264,96){\small $c\!=\!-1$}
  \put(240,25){\small $c\!=\!-1/2$}
  \put(161,96){\small $c\!=\!1/2$}
  \put(89,96){\small $c\!=\!1$}
\end{picture}
\includegraphics[scale=1]{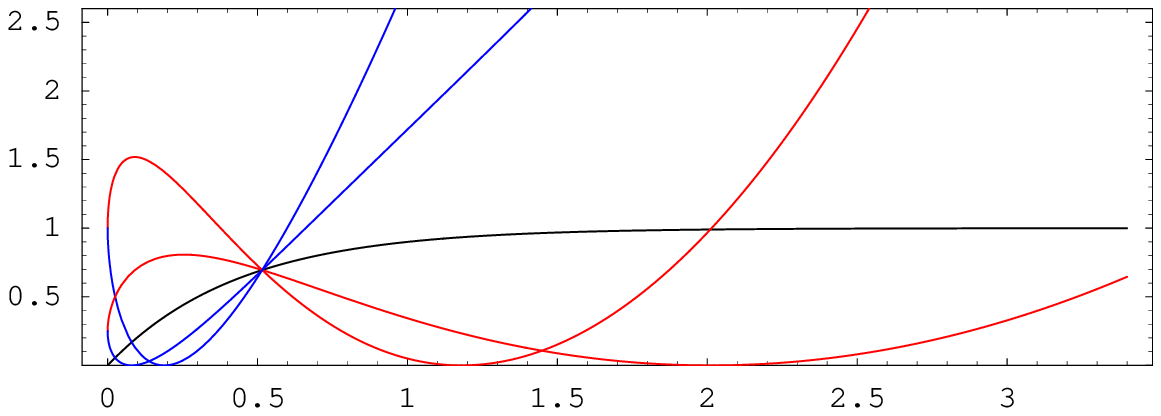}
\caption{\label{fig_bh} \small{Spacetimes with black-holes. All
curves have $a=1$ and $d=b$.}}
\end{center}
\end{figure}

If $c \neq 0$, the coefficient of $dr^2$ in the line element
remains identical as in the Schwarzschild spacetime. In this sense
the quantity $b$ may still be called the Schwarzschild radius. The
coefficient of $dt^2$ in the line element, however, is for $c\neq
0$ very different than before. This can be seen in Fig.
\ref{fig_bh} where it is plotted for a few combinations of
parameters. A number of new properties are worth pointing out.

First, the functions $f^2$ have a root at a finite radius that is
different from the usual $r=b$. Because of the dependence of $q$
on $f$, the imperfect fluid component diverges at those roots. The
curvature scalar $R_{\mu\nu\lambda\sigma}R^{\mu\nu\lambda\sigma}$
also diverges there and so these points (actually, surfaces of
spheres) are likely unphysical features of the solutions. Their
location moves close to $r=b$ as $c$ approaches zero through
positive values, and away from $r=b$ as $c$ approaches zero
through negative values. If the purpose of the spacetime is to
describe behavior outside the black hole, positive $c$ should
probably be preferred.

Second, the solutions are only valid for $r>b$ and it is not clear
whether they can be extended to $r<b$. Moreover, the quantity
$f^2$ is not zero at $r=b$ so even if the solution is continued to
smaller values of $r$ the spacetime signature may change. Although
$r=b$ is always ``hidden'' behind the singularity described above,
this is nonetheless an unphysical feature that suggests that the
spacetime should only be studied far from $r=b$.

Third, for large $r$, the functions diverge rather than approach a
constant and so they do not describe asymptotically flat
spacetimes. This occurs even though the nonvanishing component of
the stress-energy tensor, proportional to $q$, falls to zero as
$r\ra\infty.$ This is yet another undesirable feature because it
implies that the spacetime, even though it is a modification of
Schwarzschild space, does not have a limit in which it reproduces
Newtonian physics.

As a special case, it is possible to consider the solution with
$b=0$. In the Schwarzschild solution, this implies removing the
black hole from the spacetime leaving only flat space. In the case
with $c\neq 0$, setting $b=0$ gives \be f \ra a-c+c\,
\mathrm{log}\, \sqrt{\frac{4r}{d}}. \ee While this is valid for
all $r>0$, the other two features of the modified black hole
spacetimes mentioned above persist. Namely, the function has a
root at some finite value of $r$ where the stress-energy tensor
component $q$ as well as the scalar curvature
$R_{\mu\nu\lambda\sigma}R^{\mu\nu\lambda\sigma}$ diverge, and it
does not asymptote to a constant for large $r$. Thus, the
resulting spacetime represents quite a drastic departure from
Minkowski spacetime and should be interpreted similarly as above.

\subsection{Cosmological Constant \label{s_newSdS}}

The case of the cosmological constant is defined by a perfect
fluid with \be \rho=-p=\Lambda, \ee where $\Lambda$ is a nonzero
spacetime constant. The strategy for obtaining solutions in this
case is the same as before. Equation (\ref{Gtt_sc}) can first be
solved for $h$ to give \be \label{h2solution} h =
1-\frac{b}{r}-\frac{\Lambda r^2}{3} \ee with $b$ a constant. Then,
this can be substituted into (\ref{Gaa_sc}) to give a linear
equation for $f$, \be \label{eqnotsolved} 2r^2\left(3b-3r+\Lambda
r^3 \right)f^{\prime\prime} + r\left(3b-6r+4\Lambda r^3
\right)f^\prime - \left(3b+4\Lambda r^3 \right)f =0. \ee

In the special case $b=0$ and $\Lambda<0$, the solution to
(\ref{eqnotsolved}) can be found analytically and then inserted
into (\ref{Grr_sc}) to give $q$. The result is
\begin{align}
\label{fcosmo} f &= a \left(\sqrt{1-\frac{\Lambda r^2}{3}}\right)
+ \frac{c}{2} \left(1-\sqrt{1-\frac{\Lambda r^2}{3}}\,
\mathrm{arctanh}\left(\sqrt{1-\frac{\Lambda r^2}{3}} \right)^{-1}
\right), \\ \label{qcosmo} q \! &= \frac{c}{r^2 f},
\end{align} with $a$ and $c$ constants.

The effect of the imperfect fluid can again be removed by setting
$c=0$. The resultant function $f$ with $c=0$ and $a=1$ set by
convention, when inserted into the line-element ansatz, together
with (\ref{h2solution}) with $b=0$, gives anti-de Sitter
spacetime. This is actually a homogenous, rather than only
isotropic, space because $b=0$.

\begin{figure}[tbp]
\begin{center}
\begin{picture}(0,0)(0,0)
  \put(360,12){$r|\Lambda|^{1/2}$}
  \put(36,127){$f^2$}
  \put(68,59){\small $c\!=\!0$}
  \put(178,97){\small $c\!=\!-1$}
  \put(117,86){\small$c\!=\!-1/2$}
  \put(178,32){\small $c\!=\!1$}
  \put(63,37){\small $c\!=\!1/2$}
\end{picture}
\includegraphics[scale=1]{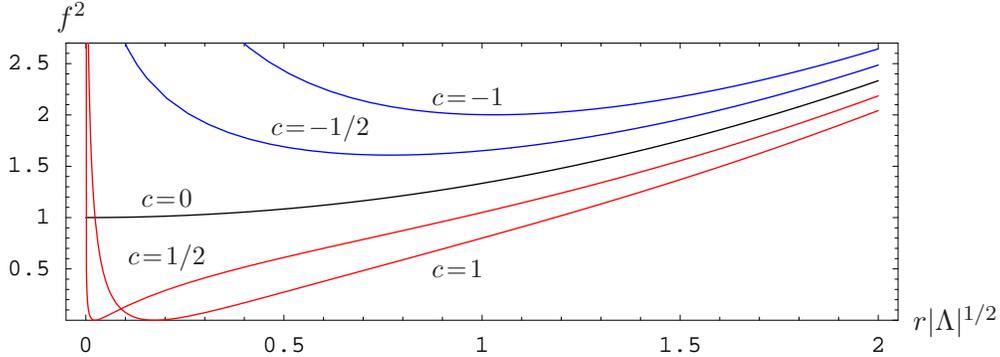}
\caption{\label{fig_ads} \small{ Spacetimes with negative
cosmological constant. All curves have $a=1$ and $\Lambda<0$. }}
\end{center}
\end{figure}

If $c\neq 0$, the solution describes a departure from anti-de
Sitter. Figure \ref{fig_ads} shows the quantity $f^2 = -g_{tt}$
for a few choices of parameters. For $c>0$, the curves have a root
at some finite radius where $q$ and the curvature invariant
$R_{\mu\nu\lambda\sigma}R^{\mu\nu\lambda\sigma}$ diverge. Curves
with $c<0$ are nonsingular everywhere. For both signs, the curves
significantly differ from the anti-de Sitter curve ($c=0$) near
$r=0$ but track it closely for $r\ra\infty$. The modified
spacetimes are thus asymptotically anti-de Sitter.

For $\Lambda>0$, it can be checked that the function \be f_{SdS} =
h^{1/2}=\sqrt{1-\frac{b}{r}-\frac{\Lambda r^2}{3}} \ee is a
solution to (\ref{eqnotsolved}) and corresponds to $q=0$. It
describes Schwarzschild-de Sitter spacetime if $b\neq 0$ and a de
Sitter spacetime if $b=0$. A feature of these spaces is that they
contain a cosmological horizon at a radius $r_c$ for which
$f_{SdS}(r_c)=0$. The metric ansatz in static coordinates suffers
from a coordinate singularity at this radius as both $g_{tt}$ and
$g_{rr}$ change sign and $g_{rr}$ diverges. Because of this
solutions must be specified separately for regions inside and
outside the cosmological horizon. While this is not a serious
hurdle when $q=0$, it does complicate matters when $q\neq 0$. For
example, (\ref{fcosmo}) and (\ref{qcosmo}) can be used for
$\Lambda>0$ and $b=0$ to describe solutions outside the horizon
but they cannot be used in the region inside the horizon because
$f^2$ is not real-valued everywhere there for any choice of
parameters $a$ and $c$. A solution inside the horizon does exist -
it can be computed numerically given some boundary data - but it
is not described by (\ref{fcosmo}) and (\ref{qcosmo}).

At this stage, it is worth noting that the function $q$ in
(\ref{qbh}) and (\ref{qcosmo}) has the same form in two
inequivalent situations. It seems reasonable therefore to guess
that this form is also valid in the more general case for
$\Lambda>0$ and $b\neq 0$. With this guess, the general solution
for $f$ can be obtained by solving equation (\ref{Grr_sc}) and
then verified by substitution into (\ref{Gaa_sc}) and
(\ref{eqnotsolved}). The guess works and the result is
\begin{align} \label{fposlambda}
f&=a\sqrt{1-\frac{b}{r}-\frac{\Lambda r^2}{3}}-c f_{c,\Lambda,b} \\
q &= \frac{c}{r^2 f} \end{align} where $f_{c,\Lambda,b}$ is a
function that can be written analytically but has a complicated
form that does not provide insightful information.

\begin{figure}[tbp]
\begin{center}
\begin{picture}(0,0)(0,0)
  \put(360,12){$\mathrm{log}_{10} r/b$}
  \put(38,127){$f^2$}
  \put(172,34){\small $c\!=\!1/4$}
  \put(277,24){\small $c\!=\!1/2$}
  \put(144,58){\small $c\!=\!0$}
  \put(258,104){\small $c\!=\!-1/2$}
  \put(106,100){\small $c\!=\!-1/4$}
\end{picture}
\includegraphics[scale=1]{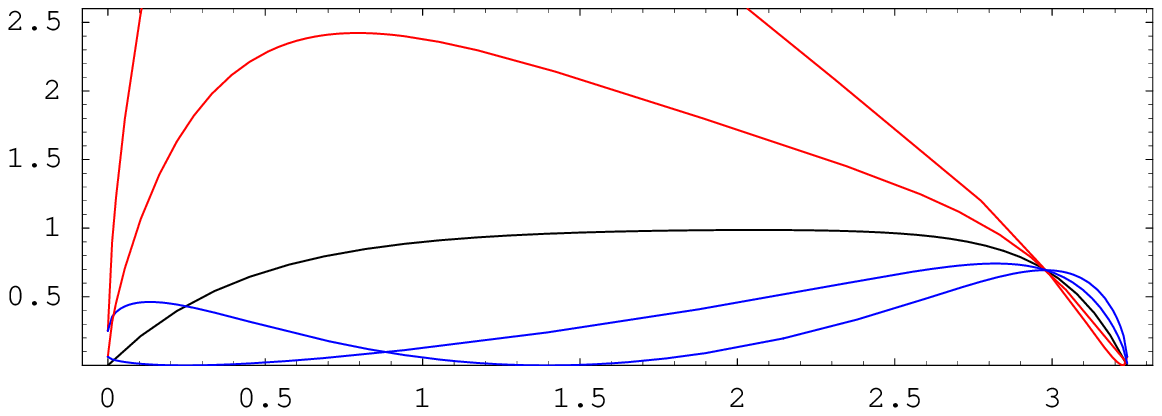}
\caption{\label{fig_SdS} \small{ Spacetimes with black-holes and
cosmological constant. All curves have $\Lambda b^2 = 10^{-6}$ and $a=1$. }}
\end{center}
\end{figure}

Fig. \ref{fig_SdS} shows $f^2 =-g_{tt}$ obtained using
(\ref{fposlambda}) with various parameters. All curves are plotted
in the domain of $r$ between the two horizons of the
Schwarzschild-de Sitter solution. Some of the features that
distinguish them from the Schwarzschild-de Sitter case are
qualitatively similar to those already found in the modified
Schwarzschild spacetimes ($\Lambda=0$).

First, all curves with $c\neq 0$ have zeros at radii that are
different from the black-hole and cosmological horizons in the
unmodified spacetime. The curvature invariant
$R_{\mu\nu\lambda\sigma}R^{\mu\nu\lambda\sigma}$ diverge at those
zeros. Their location moves close to the black-hole horizon as $c$
approaches zero through positive values, and move close to the
cosmological horizon as $c$ approaches zero through negative
values.

Second, for $c\neq 0$, the curves do not fall to zero at the black
hole or cosmological horizons. This poses problems with signature
change.

Third, the maxima of the curves with $c\neq 0$ are shifted
relative to the $c=0$ case. For $c>0$, the maximum occurs closer
to the cosmological horizon. For $c<0$, it occurs closer to the
black-hole horizon.

\subsection{Almost-Constant Perfect Fluid \label{s_fancyholes}}

This case summarizes and elaborates on the discussion in
\cite{Mannheim:2005bfa}. It involves a more qualitative
description of the effect of the imperfect fluid component on
geometry when the assumption that $\rho$ and $p$ be strict
spacetime constants is relaxed. The profiles for $\rho$, $p$ and
$q$ are assumed to satisfy
\begin{subequations} \begin{align} \label{fancyrho} \rho &=
\left\{
\begin{array}{ll} \rho(r) \;\; &\mbox{for} \;\; r< R_\rho
\\ \Lambda \;\; &\mbox{for} \;\; r\geq R_\rho
\end{array} \right. \\ \label{fancyp} p &= \left\{
\begin{array}{ll} p(r) \;\; &\mbox{for} \;\; r< R_p
\\ -\Lambda \;\; &\mbox{for} \;\; r\geq R_p
\end{array} \right.\\ \label{fancyq} q &= \left\{
\begin{array}{ll} q(r) \;\; &\mbox{for} \;\; r< R_q
\\ 0 \;\; &\mbox{for} \;\; r\geq R_q
\end{array} \right.\end{align} \end{subequations}
with appropriate continuity conditions at $r=R_\rho,R_p,R_q$. The
boundary radii define compact regions where the corresponding
stress-energy tensor components differ from constants. A
cosmological constant is allowed for large $r$ in the profiles for
$\rho$ and $p$ but not for $q$.

When (\ref{fancyrho}) is inserted into (\ref{Gtt_sc}), the
resulting equation can be solved in the region $r>R_\rho$ to give
the solution \be h = 1-\frac{b}{r} -\frac{\Lambda r^2}{3}\qquad
\mbox{for}\quad r>R_\rho. \ee This can then be inserted into
(\ref{Grr_sc}) and (\ref{Gaa_sc}). The results that ensue depend
on the relative sizes of $R_\rho$, $R_p$ and $R_q$. There are six
possible orderings and these can be treated exhaustively.

In general, the strategy for obtaining exact solutions is to solve
(\ref{Gaa_sc}) for $r>R_\rho$ and $r>R_p$ where the details of the
$\rho$ and $p$ profiles are not important. The equation, and
therefore the solution, in this region are equivalent to that
studied in the previous sections. When the solution is inserted
into (\ref{Grr_sc}), the necessary form for $q$ is obtained and it
is not a function with compact support as required by
(\ref{fancyq}). To be consistent, then, it is necessary to require
that $q=0$ in that region. Thus, the exterior solution becomes \be
\label{boringf1} f = a\sqrt{h} \qquad \mbox{for} \quad r>R_\rho,
\;\; r>R_p. \ee The pathologies associated with large radii
described for the idealized solutions in Sec. \ref{s_newhole} and
\ref{s_newSdS} are thus removed.

\begin{figure}[]
\begin{center}
\includegraphics[scale=1.0]{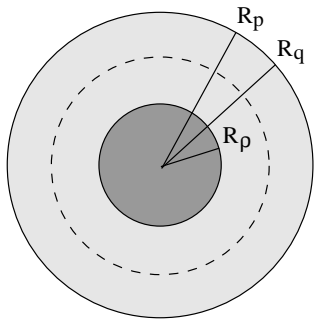}
\caption{\label{fig_compact} \small{Isotropic spacetime with
$\rho$, $p$ and $q$ profiles with compact support.}}
\end{center}
\end{figure}

An interesting case is when $R_p=R_q >R_\rho$, illustrated in Fig.
\ref{fig_compact}. As in the general case, the solution in the
exterior region $r>R_p$ reduces to (\ref{boringf1}). However, in
this case there is a shell $R_\rho < r < R_p$ where \be f \neq
a\sqrt{h}. \ee This general feature may also be argued to hold in
regions $r<R_\rho$ where $q\neq 0$ \cite{Jacobson:2007tj}.

The precise form of $f$ in the shell $R_\rho < r < R_p$ depends on
the pressure profile $p$. Under certain conditions, however, it
may be reasonable to approximate $f$ using (\ref{fbhsol}) or
(\ref{fposlambda}) when $p$ does not differ significantly from a
constant. This can only be done in a region from $r=R_\rho$ to
some intermediate radius shown by the dashed line in Fig.
\ref{fig_compact}. Beyond this intermediate radius, the pressure
must differ sufficiently from a constant in order to ensure that
$f$ goes to (\ref{boringf1}) for $r>R_p$.

\section{Phenomenology \label{s_pheno}}

The most interesting among the solutions with imperfect fluids is
the modified black hole spacetime with compact source of Sec.
\ref{s_fancyholes}. However, that solution cannot be written
explicitly without assuming profiles for $\rho$, $p$ and $q$. In
contrast, the modified black hole spacetime of Sec.
\ref{s_newhole} can be written down analytically in terms of only
two free parameters. Moreover, as argued in Sec.
\ref{s_fancyholes}, it may be a reasonable approximation to the
more physical spacetime in a region close to the matter
distribution with energy density $\rho$ and away from $r=\infty$
(or $r=r_c$ if $\Lambda >0$). So it may be instructive to consider
that idealized spacetime as a concise model of the possible
effects of a radial imperfect fluid component on geometry. The
model does have limitations, however. In particular, it does not
easily allow to estimate effects due to multiple source. Bearing
such limitations in mind, this section aims to compare the
phenomenology of the modified spacetime, following
\cite{Waldbook}, with the Schwarzschild black hole.


The line element of the modified Schwarzschild solution, repeated
here for convenience, is \be \label{ansatzrepeat} ds^2 = - f^2
\,dt^2 + h^{-1}\, dr^2 + r^2 \, d\Omega^2 \ee with \begin{align}
\label{fagain} f &= a \left(\sqrt{1-\frac{b}{r}}\right)-c\left( 1
- \sqrt{1-\frac{b}{r}} \, \mathrm{log}
\left(\frac{\sqrt{r}+\sqrt{r-b}}{\sqrt{d}} \right)\right) \\ h &=
1-\frac{b}{r}. \end{align} It reduces to that of Schwarzschild
spacetime for $a=1$ and $c=0$ and describes modification thereof
when $c\neq 0$. In the analysis below, all analytic results are
shown with $a$ explicitly present but all numerical quantities are
evaluated using a convention setting $a=1$.

\subsection{Orbital motion}

A unit four-velocity vector $u^\mu = dx^\mu/d\tau$, where $\tau$
is the proper time, should obey the normalization condition $ -1 =
u_\mu u^\mu.$ Assuming motion in the plane $\theta=\pi/2$, this
condition gives \be \label{geo2} -1 = -f^2 \, \dot{t}^2 + h^{-1}
\, \dot{r}^2 + r^2\, \dot{\phi}^2 \ee with the dot denoting $
d/d\tau.$ Because $(d/dt)^\mu$ and $(d/d\phi)^\mu$ are Killing
fields of the static isotropic spacetime, the quantities $E = f^2
\, \dot{t}$ and $L = r^2 \, \dot{\phi}$, called energy and angular
momentum, respectively, are constants of motion. Writing
(\ref{geo2}) in terms of $E$ and $L$ yields \be \label{geo3}
\frac{1}{2}\dot{r}^2 + V_{eff} = \frac{1}{2}E^2 \ee with \be
V_{eff} = \frac{h}{2}\left(1+\frac{L^2}{r^2} \right) +
\frac{1}{2}E^2 \left(1-\frac{h}{f^2} \right). \ee This is an
equation for a test particle moving in one dimension in an
effective radial potential $V_{eff}$.

A test particles moving in a circular orbit is at the minimum of
the potential and has $dV_{eff}/dr = 0.$ Since it does not move
radially, it also has $\dot{r}=0$ so that $V_{eff}=E^2/2$.
Together, these two conditions can be used to solve for $L$ and
$E$ in terms of the orbit radius. The general result
\begin{align}
E^2 &= \frac{f^3}{f-rf^\prime} \sim a^2+ac\, \mathrm{log}\frac{4r}{d} \\
L^2 &= \frac{r^3 f^\prime}{f-rf^\prime} \sim
\frac{cr^2}{2a}+\frac{br}{2}
\end{align} is independent of the function $h$. The last
approximations are obtained after expanding for large $r$ and
small $c$ and ignoring all but the leading terms that are either
proportional to $c$ or are independent of it.

The general result reproduces the Schwarzschild scalings when
$c=0$. For $c\neq 0$, however, the correction terms can dominate
for $r$ sufficiently large. In particular, the scaling of angular
momentum squared, $L^2$, shifts from linear to quadratic near  \be
\label{rtransition} r_\star = \left|\frac{ab}{c}\right|. \ee

Since angular momentum $L$ is the product of radius $r$ and
velocity $v$, a scaling $L^2 \propto r^2$ implies that the
velocity of orbiting particles is independent of distance.
Interestingly, this behavior is actually observed within galaxies
where it is interpreted either as evidence for the existence of
dark matter or as a motivation for modifying the dynamics of
general relativity. Phenomenologically, the observed behavior is
well-described by the framework of Modified Newtonian Dynamics
(MOND) \cite{Milgrom:2008rv,Sanders:2002pf} which postulates that
orbiting bodies deviate from standard Newtonian behavior and start
orbiting at a constant velocity when their acceleration falls
below the threshold $10^{-27} m^{-1}$. (And by what seems a
coincidence, the acceleration threshold is within an order of
magnitude of the root of the observed cosmological constant
$\Lambda$.)   Setting this equal to the Newtonian expression
$b/r^2$ implies that the threshold radius in the MOND description
is \be \label{rmond} r_{\star\, M\!O\!N\!D} = \left(3\!\cdot\!
10^{13}\, m^{1/2} \right) \sqrt{b} . \ee

If the derived threshold (\ref{rtransition}) is to account for
galaxy rotation curves, then, assuming a fixed parameter $a$, the
parameter $c$ should depend on $b$ as \be \label{csqrtb} c= \gamma
\sqrt{b} \ee with $\gamma$ some quantity with units of $m^{-1/2}$.
Substituting this relation into (\ref{rtransition}) and setting
that equal to (\ref{rmond}) gives \be \label{gammaMOND} \gamma_M =
3\!\cdot\! 10^{-14} \, m^{-1/2}. \ee The subscript denotes this
value is obtained from the MOND phenomenological description.

If the derived threshold (\ref{rtransition}) is said not to be
responsible for galaxy rotation curves, then a bound on $c$ can be
formulated from each galaxy's data. It is convenient to summarize
those bounds using the parameter $\gamma$ from (\ref{csqrtb}).
Since the size of a galaxy is typically an order of magnitude
larger than (\ref{rmond}), the bound on $\gamma$ would in this
case be about an order of magnitude lower than (\ref{gammaMOND}).

Within the solar system, objects orbit the Sun according to the
unmodified scaling at least up to radii $50$ times the Earth's
orbit radius. Using $b$ as the Schwarzschild radius of the Sun,
these observations yield the bounds \be \label{boundcsorbit} |c_S|
< 10^{-10}, \qquad |\gamma_S| < 10^{-13} m^{-1/2} \ee The
subscripts denote that the bound refers to an observation wherein
the central massive object is the Sun and the bound on $\gamma$ is
presented assuming the scaling (\ref{csqrtb}).

\subsection{Perihelion Precession}

If a test particle is displaced from a circular orbit, it
oscillates around the equilibrium radius with frequency $\omega_r$
given by $\omega_r^2 = d^2V_{eff}/dr^2$ with the right-hand-side
evaluated at the orbit radius. Its angular frequency is
$\omega_\phi = L/r^2$ evaluated at the orbit radius. When these
are not equal, the object precesses with frequency $\omega_p =
\omega_\phi - \omega_r.$ Using the modified black-hole spacetime
and evaluating $\omega_p$ for large $r$ and small $c$ yields \be
\omega_p \sim \frac{3\sqrt{2}\, b^{3/2}}{4\, r^{5/2}} -c
\frac{\sqrt{2}}{4\, a\sqrt{br}} \ee The precession frequency for
an elliptic but close to circular orbit differs from this result
by a factor of order one \cite{Waldbook}.

The precession of the planet Mercury around the Sun is consistent
with the $c=0$ form of $\omega_p$ with an accuracy of $10^{-5}$
\cite{Kagramanova:2006ax}. This leads to \be
\label{boundperihelion} |c_S| < 10^{-19}, \qquad |\gamma_S| <
10^{-21}\, m^{-1/2}. \ee Again, the subscripts denote that the
central source is the sun. The bound on $\gamma_S$ is obtained
using (\ref{csqrtb}).

\subsection{Redshift}

In the static coordinates of (\ref{ansatzrepeat}), $\xi^\mu =
(1,0,0,0)^T$ is a time-like Killing vector. Its magnitude is
$(-\xi_\mu \xi^\mu)^{1/2}=f$ and so a unit four-velocity vector
$u^\mu$ parallel to $\xi^\mu$ can be written as $u^\mu = f^{-1} \,
\xi^\mu.$ The frequency $\omega$ of a light ray with momentum
vector $k^\mu$ recorded by an observer with four-velocity $u^\mu$
is $\omega = k_\mu u^\mu = f^{-1} \, k_\mu \xi^\mu.$ If two
observers with unit four-velocity located at two different
locations labeled by radii $r_1$ and $r_2$ measure frequencies
$\omega_1$ and $\omega_2$ for the same light ray, the ratio of
their results is then \be \frac{\omega_1}{\omega_2} =
\frac{f(r_2)}{f(r_1)}. \ee
Without loss of generality, the coordinate $r_2$ can be replaced by $r_2 = r_1 (1+\delta)$ for some $\delta\geq-1$.

It is convenient to also define a quantity \be Y =
\frac{\omega_2-\omega_1}{\omega_2} = 1 - \frac{\omega_1}{\omega_2}
\ee that measures the relative change in the measured frequencies.
This can be computed in the limit of large $r_1$, small $c$ and
small $\delta$ to yield \be \label{Yresult} Y \sim -\frac{b
\delta}{2r}-\frac{c\delta}{2a} \ee after omitting terms of order
$c^2$, $r^{-2}$, $\delta^2$, $c/r$, and higher.


In Earth-based redshift experiments, the accuracy with which
observations match (\ref{Yresult}) with $c=0$ is
$7\!\cdot\!10^{-5}$ \cite{Vessor}. This gives \be |c_E| <
10^{-15}, \qquad |\gamma_E| < 10^{-12}\, m^{-1/2}, \ee  where the
subscripts indicate the result refers to an Earth-based
observation.

\section{Discussion \label{s_discussion}}

In a generic theory of the universe, the matter field content can
be split into standard-model fields $\psi$, a scalar field
$\varphi$, and other fields $\chi$ thought to describe dark
matter. Each of these fields can in principle give rise to a
stress-energy tensor with an imperfect fluid component $q$. Sec.
\ref{s_pheno} shows that the effects of this component in
central-mass situations can be made as small as necessary by
changing the parameter $c$. The argument from Sec.
\ref{s_fancyholes} also implies that the imperfect fluid effects
can be totally removed at large scales while producing effects
near conventional matter sources that are similar to those usually
associated with dark-matter. This section thus aims to combine
theoretical ideas and phenomenological constraints in order to
understand to what extent and under what conditions the imperfect
fluid component can account for dark-matter effects and thereby
dispense with the need to postulate the existence of separate
dark-matter fields $\chi$.

For simplicity, attention can be concentrated on the scalar field
$\varphi$. A large class of theories involving such a field have
an action of the form \be S = \int d^4x \, \sqrt{-g} \left(
\frac{1}{2} g^{\alpha\beta} \pa_\alpha \varphi \pa_\beta \varphi +
V(\varphi)\right), \ee where $V(\varphi)$ is a potential that does
not depend on derivatives of $\varphi$. The stress-energy tensor,
obtained by variation with respect to the metric, is \be
\label{Tfield} T_{\mu\nu} = \pa_\mu \varphi \pa_\nu \varphi -
\frac{1}{2} g_{\mu\nu}
 g^{\alpha \beta} \pa_\alpha \varphi \pa_\beta \varphi -
g_{\mu\nu} V(\varphi). \ee For a field configuration with $\pa_\mu
\varphi = 0$, the tensor becomes $T_{\mu\nu} = -g_{\mu\nu}
V(\varphi)$ and thus describes a cosmological constant with
$\Lambda = V(\varphi)$. On small scales, it is plausible for the
field $\varphi$ to respond to the curvature of spacetime around
massive objects, break the condition $\pa_\mu \varphi = 0$ in
certain regions, and thus lead the stress-energy tensor to acquire
more structure. In static and isotropic regions, the tensor will
take the general form (\ref{T_CMO}) with the energy density
$\rho$, pressure $p$, and imperfect fluid component $q$ related to
the derivatives of the field. In this way, the component $q$
acting like a dark field can become a companion of a conventional
energy density $\rho$ \cite{Sobouti:2008xz,Sobouti:2009pa}.

As shown in Sec. \ref{s_static}, the imperfect fluid component $q$
in a static isotropic spacetime takes the form $q\sim c/r^2 f.$
The quantity $c$ that determines its magnitude is a real
dimensionless number and it is not unreasonable for it to actually
represent a ratio of physically relevant scales. The dimensionful
quantities in the general central source setup of Sec.
\ref{s_fancyholes} are $\Lambda$, $\rho$, $p$, $R_\rho$, $R_p$,
and $R_q$ (henceforth the quantities $\rho$ and $p$ are taken to
represent averages of energy density and pressure over the
relevant regions of spacetime). The quantity $b$, the
Schwarzschild radius of the energy density distribution, is a
derived quantity proportional to $\rho R_\rho^3$. In the situation
depicted in Fig. \ref{fig_compact}, the radii obey $R_p\sim R_q >
R_q$ and it is reasonable to assume that $R_p$, $R_q$ and $p$ are
somehow related to $\Lambda$. The independent dimensionful
quantities are thus reduced to $\Lambda$, and two of $\rho$,
$R_\rho$, and $b$.

Other dimensionful parameters may enter from the specific theory
describing the field $\varphi$. In a quantum theory, for example,
a new dimensionful quantity is the Planck mass $M_P$. In an
interacting theory, dimensionful parameters may also appear from
coupling constants in potentials involving $\varphi$. For
simplicity, however, the following arguments assume that these
particle-physics parameters are not important.

The most general dimensionless quantity that can be composed of
the relevant parameters can then be written as \be
\label{generalc} c = \alpha \, b^{n_b} \, \Lambda^{n_\Lambda}\,
\rho^{n_\rho} = \beta \, b^{n_b} \, \Lambda^{n_\Lambda} \,
R_\rho^{n_R} \ee where $\alpha$ and $\beta$ are real numbers
either of order one or with a physical interpretation, and the
exponents obey \be 0= n_b-2n_\Lambda-2n_\rho = n_b-2n_\Lambda+n_R
. \ee The discussion of orbital motion in Sec. \ref{s_pheno}
suggests that the scaling should be $c=\gamma\sqrt{b}$, which
corresponds to setting $n_b=1/2$.

The value of $\gamma$ appears to be (\ref{gammaMOND}) in galaxies
but the bound (\ref{boundperihelion}) from Mercury perihelion
precession indicates that it must be significantly smaller within
the solar system. Thus, if $\gamma$ is a universal constant, the
perihelion precession observation rules out using the imperfect
fluid effect on orbital motion as a description for dark matter
leading to flat galaxy rotation curves.

If $\gamma$ is not a universal constant, then the powers $n_\rho$
or $n_R$ in (\ref{generalc}) should be different from zero. If
$n_\Lambda>1/4$ and $n_\rho < 0$, the effective value of $c$ can
be much different in galaxies and within the solar system. For
example, for $n_\Lambda=1$, $c$ becomes \be \label{cresult} c =
\alpha \, b^{1/2} \, \Lambda \, \rho^{-3/4} = \beta \, b^{1/2}\,
\Lambda \, R^{3/2}. \ee If $\rho\sim \Lambda$ and $\alpha\sim 1$,
then $c\sim b^{1/2} \Lambda^{1/4}$ and the boundary between linear
and quadratic angular momentum scaling occurs at a radius close to
(\ref{rmond}). A similar boundary radius can be obtained by taking
a larger and more realistic value for $\rho$ for a galaxy and
multiplying by a larger number $\alpha$. For a dense system, $\rho
\gg \Lambda$, the value of $c$ becomes much smaller and the
transition radius (\ref{rtransition}) correspondingly larger. In
this way, the imperfect fluid model with such scaling can be made
to account for galaxy rotation curves and also be consistent with
all the observational bounds described in Sec. \ref{s_pheno}.

In principle, the correct scaling should be computed from some
theory using (\ref{Tfield}). Such a computation would be desirable
also because it could shed light on how multiple sources affect
the strength of the imperfect fluid in realistic mass
distributions. However, this requires solving for and manipulating
the modes of the field $\varphi$ in the curved background
described in Fig. \ref{fig_compact}, and is not straight-forward.
Without the specific theory, the rather special form of the
scaling must be regarded only as a phenomenological fit.

If the interpretation of dark matter effects in terms of an
imperfect fluid related to $\varphi$ were to hold, it would have a
number of consequences. First, it would imply that direct
experimental searches for dark matter particles should give
results consistent with properties of the field $\varphi$
responsible for the background cosmological constant. Second, the
required scaling for the parameter $c$ would imply that the
threshold radius $r_\star$ should depend on the density or size of
a galaxy. This effect may be testable once the imperfect fluid is
better understood. Third, it would imply that the pressure
component of the field responsible for $q$ cannot be constant
everywhere. If it were, the imperfect fluid phenomenology would
not be localized around the central source and would be severely
inconsistent with large-scale properties of the observed universe.

Other phenomena such as gravitational lensing and structure
formation in cosmology
\cite{Sanders:2002pf,Walker:1994,Bertone:2004pz,Clowe:2006eq} may
provide more constraints when analyzed in the context of the
imperfect fluid. Distinguishing between a dark-matter field source
and an imperfect fluid component should be possible because
because the two have slightly different signatures: whereas the
former affects both the $g_{rr}$ and $g_{tt}$ components of the
metric, the imperfect fluid affects primarily $g_{tt}.$ Related
issues have been studied in the context of alternative theories of
gravity \cite{Sanders:2002pf,Walker:1994}.

\section{Conclusion \label{s_conclusion} }

Static and isotropic geometries can be consistently coupled to a
stress-energy tensor composed of a perfect fluid component plus a
radial imperfect fluid component. In cases where the perfect fluid
component vanishes exactly or describes a cosmological constant,
the solutions to the Einstein's equations are modifications of
black hole spacetimes. One of such analytic solutions, described
in Sec. \ref{s_newhole}, can be thought of as a modification to
the Schwarzschild black-hole. Its line element differs from that
of Schwarzschild space in the $dt^2$ coefficient, which in
addition to the Schwarzschild radius $b$ also contains two new
parameters: one of these, $c$, is dimensionless; the other, $d$,
has dimensions of length or mass. The spacetime contains a
curvature singularity at $r>b$ and is not asymptotically flat.
Both these properties are pathologies that can be traced to an
assumption that the pressure of the perfect fluid component of the
stress-energy tensor vanishes exactly, and as discussed in Sec.
\ref{s_fancyholes}, may be removed by relaxing this assumption.

Despite its limitations, the modified Schwarzschild spacetime can
nonetheless be useful in describing how a radial imperfect fluid
component can affect geometry. If the modified black-hole
spacetime is treated as a model for the Earth's or the Sun's
gravitational field, then, as Sec. \ref{s_pheno} shows, various
solar system observations constrain the parameter $c$.

Among the deviations from Schwarzschild phenomenology is a shift
in scaling of angular momentum $L$ of an orbiting test body from
$L^2\sim r$ to $L^2 \sim r^2$ at large $r$. Because this kind of
behavior is an observed feature in many galaxies and one of the
motivations for studies involving dark matter or modified
gravitational dynamics, Secs. \ref{s_pheno} and \ref{s_discussion}
discuss conditions under which the imperfect fluid model can be
used as a description for dark matter while still being consistent
with solar system constraints. The conclusion is that for this to
be possible the parameter $c$ must scale with various other
quantities in a rather special way. The scaling makes the setup
somewhat contrived but also falsifiable.

That a new term in the stress-energy tensor can produce new
phenomenology should not in itself be surprising. After all, it is
an additional source in the equations of motion and thus affects
geometry. The description of dark-matter effects in terms of an
imperfect fluid component, however, may be attractive in some
regards. For example, if an imperfect fluid component of the
correct magnitude were indeed produced by a scalar field $\varphi$
associated with the cosmological constant, it may help understand
dark-matter effects without the need to introduce otherwise
undetected fields or to modify gravitational interactions. From
this perspective, the results from Sec. \ref{s_discussion} suggest
that the imperfect fluid component may at least be one of several
complementary aspects to the wider subject of dark matter.
Alternatively, that discussion can be viewed as a set of
conditions that an effective theory of a dark-energy field
$\varphi$ should satisfy for it not to be at odds with
solar-system and galaxy phenomenology.

\subsection*{Acknowledgements}

I would like to thank P. Mannheim and S. McGaugh for
correspondence.


\begin{thebibliography}{99}

\bibitem{WillTest}
 C.~M.~Will, Living Rev. Relativity {\bf 9}, (2006), 3. URL (cited on 09/04/09): http://www.livingreviews.org/lrr-2006-3 .

\bibitem{Waldbook}
 R.~M.~Wald, {\em General Relativity,} CUP 1984.

\bibitem{Mannheim:2005bfa}
  P.~D.~Mannheim,
  Prog.\ Part.\ Nucl.\ Phys.\  {\bf 56}, 340 (2006)
  [arXiv:astro-ph/0505266].

\bibitem{CMO}
  D.~Cox, P.~D.~Mannheim and J.~O'Brien,
  arXiv:0903.4381 [gr-qc].

\bibitem{Mannheim:1985pp}
  P.~D.~Mannheim and D.~Kazanas,
  Gen.\ Rel.\ Grav.\  {\bf 20}, 201 (1988).


\bibitem{Konopka:2009vq}
  T.~Konopka,
  Phys. Rev. D {\bf 79}, 085012 (2009)
  [arXiv:0904.0527 [hep-th]].

\bibitem{Konopka:2009yx}
  T.~Konopka,
  arXiv:0903.4342 [gr-qc].

\bibitem{Jacobson:2007tj}
  T.~Jacobson,
  Class.\ Quant.\ Grav.\  {\bf 24}, 5717 (2007)
  [arXiv:0707.3222 [gr-qc]].


\bibitem{Milgrom:2008rv}
  M.~Milgrom,
  arXiv:0801.3133 [astro-ph].

\bibitem{Sanders:2002pf}
  R.~H.~Sanders and S.~S.~McGaugh,
  Ann.\ Rev.\ Astron.\ Astrophys.\  {\bf 40}, 263 (2002)
  [arXiv:astro-ph/0204521].

\bibitem{Kagramanova:2006ax}
  V.~Kagramanova, J.~Kunz and C.~Lammerzahl,
  Phys.\ Lett.\  B {\bf 634}, 465 (2006)
  [arXiv:gr-qc/0602002].

\bibitem{Vessor}
  R.~F.~C.~Vessor et. al., Phys.\ Rev.\ Lett.\ {\bf 45}, 2081 (1980).

\bibitem{Walker:1994}
  M.~A.~Walker, APJ {\bf 430}, 463-466 (1994).

\bibitem{Sobouti:2008xz}
  Y.~Sobouti,
  arXiv:0810.2198 [gr-qc].

\bibitem{Sobouti:2009pa}
  Y.~Sobouti, A.~H.~Zonoozi and H.~Haghi,
  arXiv:0906.0668 [gr-qc].

\bibitem{Bertone:2004pz}
  G.~Bertone, D.~Hooper and J.~Silk,
  Phys.\ Rept.\  {\bf 405}, 279 (2005)
  [arXiv:hep-ph/0404175].

\bibitem{Clowe:2006eq}
  D.~Clowe, M.~Bradac, A.~H.~Gonzalez, M.~Markevitch, S.~W.~Randall, C.~Jones and D.~Zaritsky,
  Astrophys.\ J.\  {\bf 648}, L109 (2006)
  [arXiv:astro-ph/0608407].


\end{thebibliography}
\end{document}